\documentclass[]{article} 

\usepackage{graphicx}

\begin{document} 

\title{Theory of Interacting Neural Networks}
\author{Wolfgang Kinzel\\
Institut f\"ur Theoretische Physik\\ Universit\"at W\"urzburg\\ Am
Hubland, 97074 W\"urzburg, Germany}

\maketitle

Contribution to \emph{Networks}, ed. by H.G. Schuster and S. Bornholdt,
published by Wiley VCH
 
\section{Introduction} Neural networks learn from examples. This
concept has extensively been studied using models and methods of
statistical physics \cite{Hertz,Engel}. In particular the following
scenario has been investigated: Feedforward networks are trained on
examples generated by a different network.
 
Feedforward networks classify high dimensional data, in the simplest
case by a single output bit (1/0, wrong/correct, yes/no). They are
adaptive algorithms, their parameters (= synaptic weights) are
adapting to a set of training examples, in our case a set of
input/output pairs. After the training phase, the networks have
achieved some knowledge about the rule which has generated the
examples, the network can classify input vectors which it never has
seen before, it can generalise.
 
Several mathematical models studied before use training examples which
are generated by a different neural network, called the
``teacher''. On-line training means that the ``student'', at each
training step, receives a new example from the teacher network. Each
example is used only once for training. Hence, in this case training
may be considered as dynamics of interacting neural networks: A
teacher network is sending signals (= examples) to the student network
which is stepwise changing its weights according to the received
message.
 
Mathematical methods have been developed to calculate the properties
of the dynamics of interacting networks. In the limit of large
networks one can describe the system by a differential equation for a
few ``order parameters'', which determine, for example, the
generalisation error as a function of the number of training examples
\cite{BC}.
 
In this contribution we give an overview over recent work on the
theory of interacting neural networks. The model is defined in Section
2. The typical teacher/student scenario is considered in Section 3.  A
static teacher network is presenting training examples for an adaptive
student network. In the case of multilayer networks, the student shows
a transition from a symmetric state to specialisation. Neural networks
can also generate a time series. Training on time series and
predicting it are studied in Section 4. When a network is trained on
its own output, it is interacting with itself. Such a scenario has
implications on the theory of prediction algorithms, as discussed in
Section 5. When a system of networks is trained on its minority
decisions, it may be considered as a model for competition in closed
markets, see Section 6. In Section 7 we consider two mutually
interacting networks. A novel phenomenon is observed: synchronisation
by mutual learning. In Section 8 it is shown, how this phenomenon can
be applied to cryptography: Generation of a secret key over a public
channel.

\begin{figure}[t]
\begin{center}
\includegraphics[width=.4\textwidth]{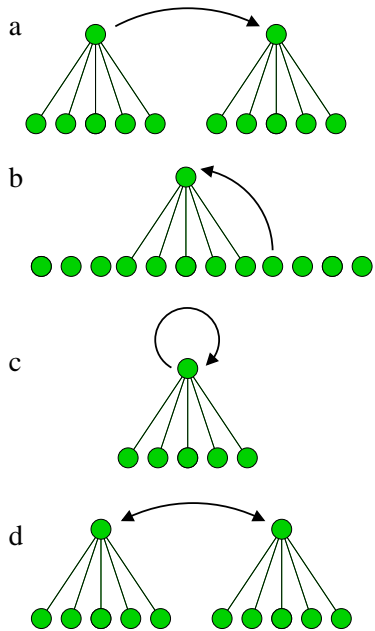}
\end{center}
\caption[]{Different kinds of interaction discussed in this contribution:
(a) static teacher vs. adaptive student, (b) time series generation and prediction, (c) self-interaction, (d) mutual learning}
\label{int}
\end{figure}

\section{On-line training} 

The simplest mathematical neural network is the perceptron. It
consists of a single layer of $N$ synaptic weights $\underline{w} =
(w_1, ..., w_N)$. For a given input vector $\underline{x}$, the output
bit is given by
 
\begin{equation}\label{one} \sigma = \mbox{sign} \left(
\sum\limits^{N}_{i=1} \; w_i x_i \right) \end{equation}
 
The decision surface of the perceptron is just a hyperplane in the
$N$-dimensional input space, $\underline{w} \cdot \underline{x} =
0$. A perceptron may also have a continuous output $y$ as
 
\begin{equation}\label{two} y = \tanh \left( \sum\limits^{N}_{i=1} \;
w_i x_i \right) \end{equation}
 
A perceptron may be considered as an elementary unit of a more complex
network like an attractor network or a multilayer network. In fact any
function can be approximated by a multilayer network if the number of
hidden units is large enough \cite{Engel}.
 
The perceptron can learn from examples. Examples are input/output
pairs,
 
\begin{equation}\label{three} (\underline{x}(t) ,
\underline{\sigma}(t)) \; \; \; \qquad t = 1,..., \alpha N
\end{equation}
 
On-line training means that at each time step $t$ the weights of the
perceptron adapt to a new example, for instance by the rule
 
\begin{equation}\label{four} \underline{w} (t+1) = \underline{w} (t) +
\frac{\eta}{N} \sigma(t) \underline{x}(t) \; F(\sigma(t)
\underline{x}(t) \cdot \underline{w}(t)) \end{equation}
 
$F(z)=1$ is usually called -- after the corresponding biological
mechanism -- the Hebbian rule, each synapse $w_i$ responds to the
activities $\sigma(t) x_i(t)$ at its ends. $F(z) = \Theta(-z)$ is
called the Rosenblatt rule: a training step occurs only if the example
is misclassified. Finally, the Adatron rule $F(z) = |z| \Theta(-z)$ is
important since it gives good results for generalisation, as discussed
in the following. For the last two learning rules, in addition to the
two neural activities at the synaptic ends, the postsynaptic potential
determines the strength of the synaptic adaptation.

\section{Generalisation}

Now we consider two perceptrons. One is called the teacher network
which is producing a set of examples. It is receiving a set of random
input vectors $\underline{x}(t)$ and generating output bits
$\sigma(t)$.  The teacher has a fixed weight vector $\underline{w}^T$.

Each time the teacher is producing a new example, the student
perceptron is trained on it according to (\ref{four}). As a consequence, the
weight vector of the student $\underline{w}^S(t)$ is time dependent.
The student tries to approach the teacher, at each training step $t$
its weight vector moves towards the one of the teacher. It is easy to
see that for random inputs $\underline{x}$, equation (\ref{four})
gives a kind of random walk in $N$-dimensional space with a bias
towards the teacher vector $\underline{w}^T$.

The distance between student and teacher can be measured by the
overlap

\begin{equation}
R=\frac{\underline{w}^T \cdot \underline{w}^S} {|\underline{w}^T|
|\underline{w}^S|}
\end{equation}
The quantity $R$ determines the angle $\phi$ between student and
teacher weights, $R=\cos\phi$. It turns out that from this overlap the
generalisation error $e_g$ can be calculated. The generalisation error
is the probability that the student gives an answer to a random input
$\underline{x}$ which is different from the one of the teacher.  One
finds

\begin{equation}
e_g = \frac{\arccos R}{\pi}
\end{equation}

In the limit of infinitely many input units, $N \to \infty$, the
dynamics of the overlap $R(t)$ can be calculated analytically.
According to (\ref{four}), the size of the training step scales down
with $1/N$. Therefore one defines a variable $\alpha=t/N$ which
becomes a continuous variable, called time, in the limit of large $N$.

The time dependence of $R(\alpha)$ is obtained by multiplying
(\ref{four}) by $\underline{w}^T$ and $\underline{w}^S$ and by
averaging these two equations over the random input vector
$\underline{x}$. This can be done since the expressions $\underline{w}
\cdot
\underline{x}$ are  Gaussian variables.

For the Adatron rule one finally obtains the differential equation
\cite{BR}

\begin{equation}
\frac{d R}{d \alpha} = - \frac{R}{2 \pi} \arccos{R} + \frac{1}{\pi}
\big(1-\frac{R^2}{2}\big)\sqrt{1-R^2}
\end{equation}

\begin{figure}[ht]
\begin{center}
\includegraphics[width=.6\textwidth]{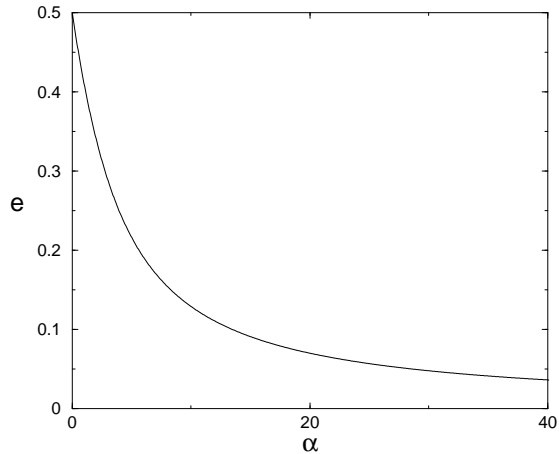}
\end{center}
\caption[]{Generalisation error as a function of time, from Ref. \cite{BR}}
\label{adatron}
\end{figure}

The generalisation error calculated by this equation is shown in
Fig. \ref{adatron}.  If only a finite number of examples has been
learned, $\alpha=0$, the error is 50 \%, as bad as random guessing.
If the number of training examples is of the order of $N$, $\alpha >
0$, the student network has obtained some overlap to the teacher.  In
the limit of large values of $\alpha$ the error decreases to zero, the
student has obtained complete knowledge about the parameters of the
teacher.

\begin{figure}[ht]
\begin{center}
\includegraphics[width=.6\textwidth]{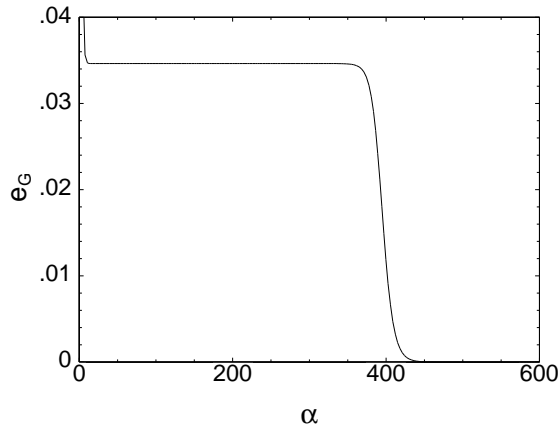}
\end{center}
\caption[]{Generalisation error as a function of time,
for a multilayer network with two hidden units, from Ref. \cite{BRW}}
\label{plateau}
\end{figure}

The asymptotic decay of the generalisation error depends on the
learning rule.  One finds for the Adatron rule $e_g \propto 3/2
\, \alpha^{-1}$. In fact, it has been shown
that the error cannot decay faster than $e_g \propto 0.88 \,
\alpha^{-1}$\cite{KC}. For the Rosenblatt rule, one finds $e_g \propto
\alpha^{-1/3}$, and for the Hebbian rule $e_g \propto
\alpha^{-1/2}$. But in all cases the student succeeds to approach the
teacher when it makes of the order of $N$ steps.  This even holds when
the examples are distorted by noise \cite{BRS}.

Learning from examples works for more complex networks, too. Here I
would like to mention the work on specialisation of committee machines
\cite{BRW}.
Such a network is a multilayer network with several hidden units,
similar to Fig. \ref{par}. The output bits of the continuous hidden units 
are summed and taken as the output of the network. Teacher and student
networks have an identical architecture, and the learning step is just
a gradient descent of the training error, the quadratic deviation
between teacher and student output.

The corresponding generalisation error is shown in Fig. \ref{plateau}. For
small number of examples it decreases fast, then it reaches a plateau
and only for a huge number of examples it decreases to zero.

The motion of the student network can be expressed by the overlap
between the corresponding members of the two machines. The teacher as
well as the student consists of two weights vectors,
$\underline{w}^{T/S}_1$ and $\underline{w}^{T/S}_2$.  A distance
between teacher and student can be defined from the overlaps

\begin{equation}
R_{i,j}=\underline{w}^T_i \cdot \underline{w}^S_j
\end{equation}
 
Initially, all vectors are random, hence up to fluctuations all the
overlaps are zero. Then all overlaps increase due to learning. But on
the plateau of Fig. \ref{plateau} the overlaps are all identical,
$R_{1,1}=R_{2,2}=R_{1,2}$. The student has achieved some knowledge,
but it is in a symmetric state. Only if the student receives much more
information it can specialise: $R_{1,1}=R_{2,2}$ are much larger than
$R_{1,2}$.

In the initial process the two members of the student committee act as
being one single perceptron, but later they specialise and follow
their partners in the teacher committee until they achieve complete
knowledge for $\alpha \to \infty$.

\section{Time series prediction and generation}

\begin{figure}[ht]
\begin{center}
\includegraphics[width=.6\textwidth]{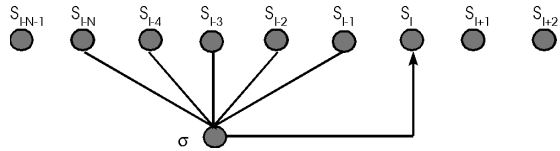}
\end{center}
\caption[]{Time series generation and prediction by a perceptron}
\label{bg}
\end{figure}

Neural networks are successful prediction algorithms \cite{WG}. Given
a sequence of numbers, a neural network can be trained on this
sequence by moving it over the sequence, as shown in Fig.\ref{bg}. This
sequence can be produced by another network, called the teacher, by
generating a new number and using it as an input component in the next
step.  Hence we have a new kind of interacting networks: The teacher
with its static weight vector is a bit or sequence generator
\cite{EKKK}.  The student is adapting its weights to the sequence
generated by the teacher.

In principle this is the scenario described in the previous section.
Here the difference is that the patterns are correlated, the input is
not random but generated from the output of the teacher.

If a neural network cannot generate a given sequence of numbers, it
cannot predict it with zero error.  Hence one has investigated the
generation of time series by neural networks \cite{KKPE,PK,EK,SK} But
this is not the whole story.  Even if the sequence has been generated
by an (unknown) neural network (the teacher), a different network (the
student) can try to learn and to predict this sequence. In this
context we are interested in two questions:

\begin{enumerate}
\item When a student network with the identical architecture as the
  teacher's is trained on the sequence, how does the overlap between
  student and teacher develop with the number of training examples (=
  windows of the sequence)?
\item After the student network has been trained on a 
part of the sequence, how well can it predict the sequence several
steps ahead?
\end{enumerate}

Recently these questions have been investigated numerically for the
simple perceptron, equation (\ref{one},\ref{two})
\cite{Freking}. 
Consider a teacher perceptron with weight vector $\underline{w}^T$
generating the sequence $S_0,S_1,S_2,...,S_t,...$ This sequence
follows the equation
\begin{equation}
S_t = f\left( \frac{1}{N} \sum\limits^{N}_{j = 1} \; w_j S_{t-j}
\right)
\end{equation}
where $f(x)$ is the transfer function.  It has been shown that a
perceptron can generate simple as well as complex sequences \cite{KKPE,PK}.

If $f(x)$ is monotonic, for instance $f(x)=\tanh(\beta x)$, then in
general one obtains quasiperiodic sequences.  In fact, the sequence is
essentially generated by one Fourier component of the weight vector
$w^T_i$
\cite{KKPE}. If the transfer function, however, is not monotonic, for
instance $f(x)=\sin(\beta x)$, then the sequence can be chaotic,
depending on the model parameters \cite{PK}. For both cases, learning
and prediction have been investigated \cite{Freking}.

If a quasi periodic sequence is learned on-line, using gradient
descent to update the weights,
\begin{equation}
\label{ten}
\Delta w_i = \frac{\eta}{N} (S_t - f(h)) \cdot f^{\prime} (h) \cdot
S_{t-i} \; \; \mbox{with} \; \; h= \beta \sum\limits^{N}_{j=1} w_j
S_{t-j}
\end{equation} 
then one has found two time scales (time $\alpha$ means the number
of training steps divided by $N$):
\begin{enumerate}
\item A short scale on which the overlap $R(\alpha)$ between teacher and
  student rapidly increases to a value which is still far away from
  the value $R=1$, which corresponds to perfect agreement.
\item A long one on which the overlap $R(\alpha)$ increases very slowly. 
Numerical simulations up to $10^6 N$ training steps yielded an overlap
which was close but sell different from the value $R=1$.
\end{enumerate}

Although there is a mathematical theorem on stochastic optimisation
which seems to guarantee convergence to perfect success
\cite{StochOpt}, the on--line algorithm cannot gain much information
about the teacher network, at least during reasonable training
periods.

This is completely different for a chaotic time series generated by a
corresponding teacher network with $f(x)=\sin(x)$. It turns out that
the chaotic series appears like a random one: After a number of
training steps of the order of $N$ the overlap relaxes exponentially
fast to perfect agreement between teacher and student.

Hence, after training the perceptron with a number of examples of the
order of $N$ we obtain the two cases: For a quasi periodic sequence
the student has not obtained much information about the teacher, while
for a chaotic sequence the student's weight vector comes close to the
one of the teacher. One important question remains: How well can the
student predict the time series?

\begin{figure}[ht]
\begin{center}
\includegraphics[width=.6\textwidth]{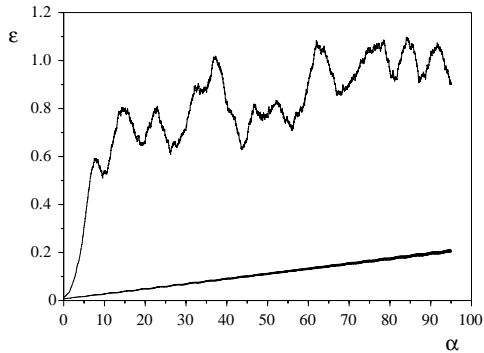}
\end{center}
\caption[]{Prediction error as a function of 
time steps ahead, for a quasi periodic (lower) and chaotic (upper)
series, from Ref.\cite{Freking}}
\label{predict}
\end{figure}

Fig.\ref{predict} shows the prediction error as a function of the time
interval over which the student makes the predictions. The student
network which has been trained on the quasi periodic sequence can
predict it very well. The error increases linearly with the size of
the interval, even predicting $10N$ steps ahead yields an error of
about 10\% of the total possible range. On the other side, the student
trained on the chaotic sequence cannot make predictions. The
prediction error increases exponentially with time; already after a
few steps the error corresponds to random guessing, $ e_g
\simeq 1$. The explanation is that an infinitesimal change in the
parameters of a chaotic map has the same effect as a small change in
initial conditions, namely, an exponential growth in the distance
between original and the disturbed trajectory.

In summary one finds the counterintuitive result:
\begin{enumerate}
\item A network trained on a quasiperiodic sequence does not 
obtain much information about the teacher network which generated the
sequence. But the network can predict this sequence over many (of the
order of $N$) steps ahead.
\item A network trained on a chaotic sequence obtains almost 
complete knowledge about the teacher network. But this network cannot
make reasonable long-term predictions on the sequence.
\end{enumerate}

It would be interesting to find out whether this result also holds for
other prediction algorithms, such as multi-layer networks.

\section{Self-interaction}

In the previous section the time series was generated by a static
teacher network. Now we consider a network which changes its synaptic
weights while it is generating a bit sequence. The teacher is
interacting with itself. The motivation of this investigation stems
from the following problem:

Consider some arbitrary prediction algorithm. It may contain all the
knowledge of mankind, many experts may have developed it. Now there is
a bit sequence $S_1, S_2, \dots$ and the algorithm has been trained on
the first $t$ bits $S_1,\dots, S_t$. Can it predict the next bit
$S_{t+1}$? Is the prediction error, averaged over a large $t$
interval, less than 50\%?

If the bit sequence is random then every algorithm will give a
prediction error of 50\%. But if there are some correlations in the
sequence then a clever algorithm should be able to reduce this
error. In fact, for the most powerful algorithm one is tempted to say
that for {\it any} sequence it should perform better than 50\%
error. However, this is not true \cite{Zhu}.  To see this just
generate a sequence $S_1, S_2, S_3,\dots$ using the following
algorithm: \\

\centerline{\fbox{
\parbox{10cm}{
Define $S_{t+i} $ to be the opposite of the prediction of this
algorithm which has been trained on $S_1, \dots, S_t$ .}}}

\vspace{4mm}
Now, if the same algorithm is trained on this sequence, it will always
predict the following bit with 100\% error. Hence there is no general
prediction machine; to be successful the algorithm needs some
pre-knowledge about the class of problems it is applied to.

The Boolean perceptron is a very simple prediction algorithm for a bit
sequence, in particular with the Hebbian on--line training algorithm
(\ref{four}).  What does the bit sequence look like for which the
perceptron fails completely?

Following (\ref{four}), we just have to take the negative value
\begin{equation}
\label{eleven}
S_t = - \mbox{sign} \left( \sum\limits^N_{j=1} \; w_j S_{t-j} \right)
\end{equation}
and then train the network on this new bit:
\begin{equation}
\label{twelve}
\Delta w_j = + \frac{1}{N} S_t \; S_{t-j}.
\end{equation}
The perceptron is trained on the opposite (= negative) of its own
prediction. Starting from (say) random initial states $S_1, \dots,
S_N$ and weights $\underline{w}$, this procedure generates a sequence
of bits $S_1, S_2, \dots S_t, \dots$ and of vectors $\underline{w},
\underline{w}(1), \underline{w}(2), \dots \underline{w}(t), \dots$ as
well. Given this sequence and the same initial state, the perceptron
which is trained on it yields a prediction error of 100\%.

It turns out that this simple algorithm produces a rather complex bit
sequence which comes close to a random one \cite{Metzler}. After a
transient time the weight vector $\underline{w}(t)$ seems to perform a
kind of random walk on an $N$--dimensional hyper-sphere. The bit
sequence runs to a cycle whose average length $L$ scales exponentially
with $N$,
\begin{equation}
\label{thirteen}
L \simeq 2.2^N.
\end{equation}
The autocorrelation function of the sequence shows complex properties:
It is close to zero up to $N$, oscillates between $N$ and $3N$ and it
is similar to random noise for larger distances. Its entropy is
smaller than the one of a random sequence since the frequency of some
patterns is suppressed. Of course, it is not random since the
prediction error is 100\% instead of 50\% for a random bit sequence.

When a second perceptron (=student) with different initial state
$\underline{w}^S$ is trained on such a anti-predictable sequence
generated by Eq.(\ref{eleven}) it can perform somewhat better than the
teacher: The prediction error goes down to about 78\% but it is still
larger than 50\% for random guessing. Related to this, the student
obtains knowledge about the teacher: The angle between the two weight
vectors relaxes to about 45 degrees \cite{Zhu,Metzler}.  Hence the
complex anti-predictable sequence still contains enough information for
the student to follow the time dependent teacher.

\section{Agents competing in a closed market}

We just considered a network interacting with itself. Now we extend
this model to a system of many networks interacting with the minority
decision of all members. This work was motivated by the following
problem of econophysics \cite{Econophys}.

Recently a mathematical model of economy receives a lot of attention
in the community of statistical physics.  It is a simple model of a
closed market: There are $K$ agents who have to make a binary decision
$\sigma(t) \in \{+1, -1\}$ at each time step. All of the agents who
belong to the minority gain one point, the majority has to pay one
point (to a cashier which always wins).  The global loss is given by
\begin{equation}
\label{fourteen}
G = \left| \sum\limits^{K}_{t=1} \; \; \sigma(t) \right|
\end{equation}
If the agents come to an agreement before they make a new decision, it
is easy to minimise $G: (K-1)/2$ agents have to choose $+1$, then $G
=1$. However, this is not the rule of the game; the agents are not
allowed to make contracts, and communicate only through the global sum
of decisions. Each agent knows only the history of the minority
decision, $S_1, S_2, S_3, \dots$, but otherwise he/she has no
information. Can the agent find an algorithm to maximise his/her
profit?

If each agent makes a random decision, then $\langle G^2
\rangle = K$. It is possible, but not trivial, to find algorithms which 
perform better than random \cite{Challet}.

Here we use a perceptron for each agent to make a decision based on
the past $N$ steps $\underline{S} = (S_{t-N}, ..., S_{t-1})$ of the
minority decision. The decision of agent $\underline{w}(t)$ is given
by
\begin{equation}
\label{fifteen}
\sigma(t) = \mbox{sign} (\underline{w}(t) \; \underline{S}).
\end{equation}
After the bit $S_t$ of the minority has been determined, each
perceptron is trained on this new example $(\underline{S}, S_t)$,
\begin{equation}
\label{sixteen}
\Delta \underline{w}(t) = \frac{\eta}{N} \; \; S_t \; \underline{S}.
\end{equation}
This problem could be solved analytically \cite{MKK}.  The average
global loss for $\eta \rightarrow 0$ is given by
\begin{equation}
\label{seventeen}
\langle G^2 \rangle = (1-2/\pi) K \simeq 0.363 \; K.
\end{equation}
Hence, for small enough learning rates the system of interacting
neural networks performs better than random decisions.  Successful
cooperation emerges in a pool of adaptive perceptrons.

\section{Synchronisation by mutual learning}

Before, we have considered a pool of several neural networks interacting
through their minority decisions.  Now we study the interaction of
just two neural networks \cite{MKK}.  Contrary to the teacher/student
case, now both of the networks are adaptive, each network is learning
the output bit of its partner.

\begin{figure}[ht]
\begin{center}
\includegraphics[width=.6\textwidth]{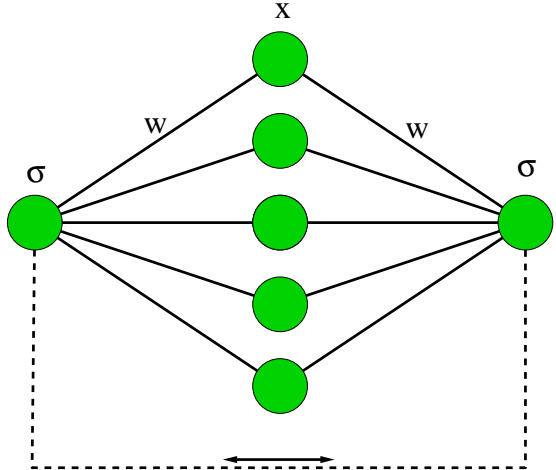}
\end{center}
\caption[]{Two perceptrons are trained by their mutual output bits, from Ref.
\cite{MKK}}
\label{per}
\end{figure}

In particular, we consider the model where two perceptrons $A$ and $B$
receive a common random input vector $\underline{x}$ and change their
weights $\underline {w}$ according to their mutual bit $\sigma$, as
sketched in Fig. \ref{per}. The output bit $\sigma$ of a single
perceptron is given by the equation
 
\begin{equation}\label{eightteen} 
\sigma = \mbox{sign} (\underline{w} \cdot \underline{x}) 
\end{equation} 
$\underline{x}$ is an $N$-dimensional input vector with components
which are drawn from a Gaussian with mean $0$ and variance
$1$. $\underline{w}$ is a $N$-dimensional weight vector with
continuous components which are normalised,
 
\begin{equation}\label{nineteen} 
\underline{w} \cdot \underline{w}= 1 
\end{equation}

The initial state is a random choice of the components $w_i^{A/B},
i=1, ... ,  N$ for the two weight vectors $\underline{w}^{A}$ and
$\underline{w}^{B}$.  At each training step a common random input
vector is presented to the two networks which generate two output bits
$\sigma^{A}$ and $\sigma^B$ according to (\ref{eightteen}). Now the weight
vectors are updated by the Rosenblatt learning rule (\ref{four}):
 
\begin{eqnarray}\label{drei} 
\underline{w}^{A} (t+1) & = & \underline{w}^{A} (t) + \frac{\eta}{N} \underline{x}\; \sigma^B \; \Theta(-\sigma^A \sigma^B) \nonumber\\ 
\underline{w}^B (t+1) & = & \underline{w}^B (t) + \frac{\eta}{N} \underline{x} \; \sigma^{A} \; \Theta (-\sigma^{A} \sigma^B) 
\end{eqnarray} 
$\Theta(x)$ is the step function. Hence, only if the two perceptrons
disagree a training step is performed with a learning rate
$\eta$. After each step (\ref{drei}), the two weight vectors have to
be normalised.
 
In the limit $N \rightarrow \infty$, the overlap
 
\begin{equation}\label{vier} 
R(t) = \underline{w}^{A} (t) \; \underline{w}^{B} (t)
\end{equation} 
has been calculated analytically \cite{MKK}.  The number of training
steps $t$ is scaled as $\alpha = t/N$, and $R(\alpha)$ follows the
equation
 
\begin{equation}\label{fuenf} 
\frac{d R}{d \alpha} = (R+1) \left( \sqrt{\frac{2}{\pi}} \; \eta(1-R) - \eta^2 \frac{\phi}{\pi} \right) 
\end{equation} 
where $\phi$ is the angle between the two weight vectors
$\underline{w}^{A}$ and $\underline{w}^B$, i.e. $R= \cos \phi$. This
equation has fixed points $R=1, R=-1$, and
 
\begin{equation}\label{sechs} 
\frac{\eta}{\sqrt{2 \pi}} = \frac{1- \cos \phi}{\phi} 
\end{equation} 
 
\begin{figure} 
\centering 
\includegraphics[width=.6\textwidth]{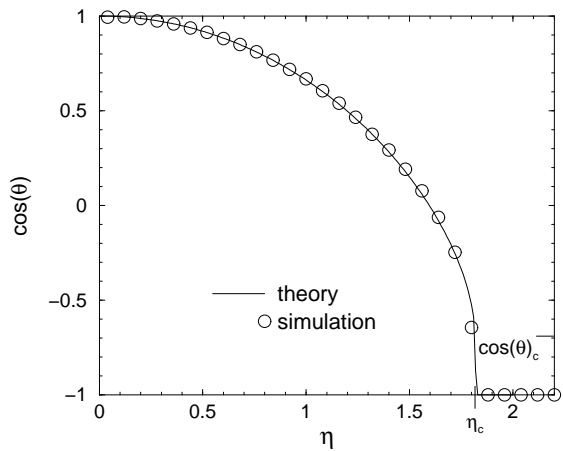} 
\caption[]{ 
  Final overlap $R$ between two perceptrons as a function of learning
  rate $\eta$. Above a critical rate $\eta_c$ the time dependent
  networks are synchronised. From Ref. \cite{MKK} }
\label{fig1} 
\end{figure} 

Fig.\ref{fig1} shows the attractive fixed point of (\ref{fuenf}) as a
function of the learning rate $\eta$. For small values of $\eta$ the
two networks relax to a state of a mutual agreement, $R \rightarrow 1$
for $\eta \rightarrow 0$. With increasing learning rate $\eta$ the
angle between the two weight vectors increases up to the value $\phi =
133^{\circ}$ for
 
\begin{equation}\label{sieben} 
\eta \rightarrow \eta_c \cong 1.816 
\end{equation} 
Above the critical rate $\eta_c$ the networks relax to a state of
complete disagreement, $\phi = 180^{\circ}, R= - 1$. The two weight
vectors are antiparallel to each other, $\underline{w}^{A} = -
\underline{w}^B$. 
 
As a consequence, the analytic solution shows, well supported by
numerical simulations for $N = 100$, that two neural networks can
synchronise to each other by mutual learning. Both of the networks are
trained to the examples generated by their partner and finally obtain
an antiparallel alignment.  Even after synchronisation the networks
keep moving, the motion is a kind of random walk on an N-dimensional
hypersphere producing a rather complex bit sequence of output bits
$\sigma ^{A} = - \sigma^B$ \cite{Metzler}. In fact, after
synchronisation the system is identical to the single network learning
its opposite output bit discussed in section 5.

\section{Cryptography}

In the field of cryptography, one is interested in methods to transmit
secret messages between two partners A and B. An opponent E who is
able to listen to the communication should not be able to recover the
secret message.
 
Before 1976, all cryptographic methods had to rely on secret keys for
encryption which were transmitted between A and B over a secret
channel not accessible to any opponent. Such a common secret key can
be used, for example, as a seed for a random bit generator by which
the bit sequence of the message is added (modulo 2).
 
In 1976, however, Diffie and Hellmann found that a common secret key
could be created over a public channel accessible to any
opponent. This method is based on number theory: Given limited
computer power, it is not possible to calculate the discrete logarithm
of sufficiently large numbers \cite{St}.
 
Recently, it has been shown how interacting neural networks can
produce a common secret key by exchanging bits over a public channel
and by learning from each other
\cite{KKK}. 

We want to apply synchronisation of neural networks to
cryptography. In the previous section we have seen that the weight
vectors of two perceptrons learning from each other can
synchronise. The new idea is to use the common weights
$\underline{w}^{A} = - \underline{w}^B$ as a key for encryption.  But
two problems have to be solved yet: (i) Can an external observer,
recording the exchange of bits, calculate the final $\underline{w}^{A}
(t)$, (ii) does this phenomenon exist for discrete weights? Point (i)
is essential for cryptography, it will be discussed further
below. Point (ii) is important for practical solutions since
communication is usually based on bit sequences.  It will be
investigated in the following.

Synchronisation occurs for normalised weights, unnormalised ones do
not synchronise \cite{MKK}. Therefore, for discrete weights, we
introduce a restriction in the space of possible vectors and limit the
components $w_i^{A/B}$ to $2L+1$ different values,
 
\begin{equation}\label{acht} 
w_i^{A/B} \in \{ -L, -L +1, ... , L-1, L \}
\end{equation} 
In order to obtain synchronisation to a parallel -- instead of an
antiparallel -- state $\underline{w}^{A} = \underline{w}^B$, we modify
the learning rule (\ref{drei}) to:
 
\begin{eqnarray}\label{neun} 
\underline{w}^{A} (t+1) & = & \underline{w}^{A} (t) - \underline{x} \sigma^{A} \Theta (\sigma^{A} \sigma^B) \nonumber \\ 
\underline{w}^B (t+1) & = & \underline{w}^B (t) - \underline{x} \sigma^B \Theta (\sigma^{A} \sigma^B) 
\end{eqnarray} 
Now the components of the random input vector $\underline{x}$ are
binary $x_i \in \{+1, -1\}$. If the two networks produce an identical
output bit $\sigma^{A} = \sigma^B$, then their weights move one step
in the direction of $-x_i \sigma^A$. But the weights should remain in
the interval (\ref{acht}), therefore if any component moves out of
this interval, it is set back to the boundary $w_i = \pm L$.
 
Each component of the weight vectors performs a kind of random walk
with reflecting boundary. Two corresponding components $w_i^{A}$ and
$w_i^B$ receive the same random number $\pm 1$. After each hit at the
boundary the distance $|w_i^{A} - w_i^B|$ is reduced until it has
reached zero.  For two perceptrons with a $N$-dimensional weight space
we have two ensembles of $N$ random walks on the internal $\{ -L, ...,
L\}$. If we neglect the global signal $\sigma^{A} = \sigma^B$ as well
as the bias $\sigma^A$, we expect that after some characteristic time
scale $\tau =
\mathcal{O}(L^2)$ the probability of two random walks being in different states 
decreases as
 
\begin{equation}\label{zehn} 
P(t) \sim P(0) e^{-t/\tau}
\end{equation} 
Hence the total synchronisation time should be given by $N \cdot P(t)
\simeq 1$ which gives 
 
\begin{equation}\label{elf} 
t_{\mathrm{sync}} \sim \tau \ln N
\end{equation} 
In fact, the simulations for $N = 100$ show that two perceptrons with
$L=3$ synchronise in about 100 time steps and the synchronisation time
increases logarithmically with $N$. However, the simulations also
showed that an opponent, recording the sequence of $(\sigma^{A} ,
\sigma^B,
\underline{x})_t$ is able to synchronise, too. Therefore, a single 
perceptron does not allow a generation of a secret key.

Obviously, a single perceptron transmits too much information. An
opponent, who knows the set of input/output pairs, can derive the
weights of the two partners after synchronisation. Therefore, one has
to hide so much information, that the opponent cannot calculate the
weights, but on the other side one has to transmit enough information
that the two partners can synchronise.
 
\begin{figure} 
\centering 
\includegraphics[width=.6\textwidth]{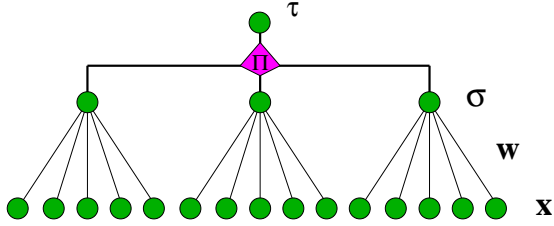} 
\caption[]{Parity machine with three hidden units.} 
\label{par} 
\end{figure} 

In fact, it was shown that multilayer networks with hidden units may
be candidates for such a task \cite{KKK}. More precisely, we consider
parity machines with three hidden units as shown in
Fig.\ref{par}. Each hidden unit is a perceptron (\ref{one}) with
discrete weights (\ref{acht}). The output bit $\tau$ of the total
network is the product of the three bits of the hidden units
 
\begin{eqnarray}\label{zwoelf} 
\tau^{A} & = & \sigma^{A}_{1} \; \sigma^{A}_{2} \; \sigma^{A}_{3} \nonumber \\ 
\tau^B & = & \sigma^{B}_{1} \; \sigma^{B}_{2} \; \sigma^{B}_{3} 
\end{eqnarray} 
At each training step the two machines $A$ and $B$ receive identical
input vectors $\underline{x}_1 , \underline{x}_2 ,
\underline{x}_3$. The training algorithm is the following: Only if the
two output bits are identical, $\tau^{A} = \tau^B$, the weights can be
changed. In this case, only the hidden unit $\sigma_{i}$ which is
identical to $\tau$ changes its weights using the Hebbian rule
 
\begin{equation}\label{dreizehn} 
\underline{w}^{A}_{i} (t+1) = \underline{w}^{A}_{i} (t) - \underline{x}_i \tau^{A} 
\end{equation} 
For example, if $\tau^{A} = \tau^B = 1$ there are four possible
configurations of the hidden units in each network:
 
\centerline{ $(+1, +1, +1), (+1,-1, -1), (-1, +1, -1), (-1, -1, +1)$} 
\noindent 
In the first case, all three weight vectors $\underline{w}_i,
\underline{w}_2, \underline{w}_3$ are changed, in all other three 
cases only one weight vector is changed. The partner as well as any
opponent does not know which one of the weight vectors is updated.
 
The partners $A$ and $B$ react to their mutual stop and move signals
$\tau^{A}$ and $\tau^B$, whereas an opponent can only receive these
signals but not influence the partners with its own output bit.  This
is the essential mechanism which allows synchronisation but prohibits
learning.  Numerical \cite{KKK} as well as analytical \cite{RKK}
calculations of the dynamic process show that the partners can
synchronise in a short time whereas an opponent needs a much longer
time to lock into the partners.
 
This observation holds for an observer who uses the same algorithm
(\ref{dreizehn}) as the two partners $A$ and $B$. Note that the
observer knows 1. the algorithm of $A$ and $B$,\, 2. the input vectors
$\underline{x}_1, \underline{x}_2, \underline{x}_3$ at each time step
and \, 3. the output bits $\tau^{A}$ and $\tau^B$ at each time step.
Nevertheless, it does not succeed in synchronising with $A$ and $B$
within the communication period.
 
\begin{figure} 
\centering 
\includegraphics[width=.6\textwidth]{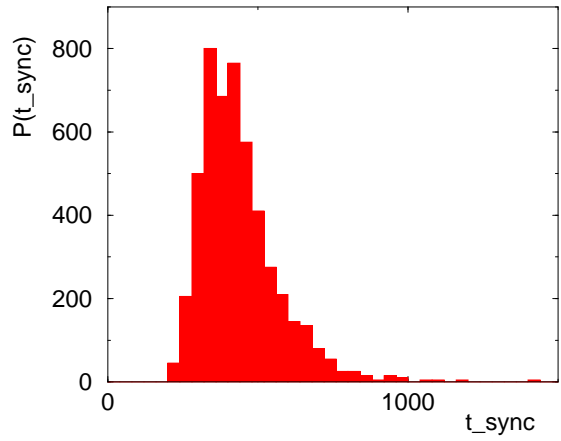} 
\caption[]{Distribution of synchronisation time for $N=100, L=3$, 
from Ref.\cite{KKK}} 
\label{tsync} 
\end{figure}

\begin{figure} 
\centering 
\includegraphics[width=.6\textwidth]{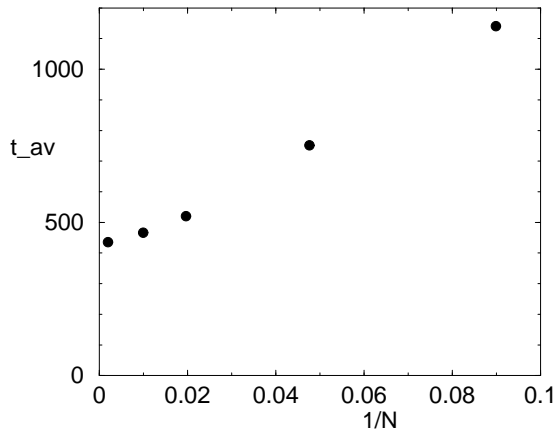} 
\caption[]{Average synchronisation time as a function of inverse system size.} 
\label{tn} 
\end{figure} 

Since for each run the two partners draw random initial weights and
since the input vectors are random, one obtains a distribution of
synchronisation times as shown in Fig. \ref{tsync} for $N=100$ and
$L=3$.  The mean value of this distribution is shown as a function
of system size $N$ in Fig. \ref{tn}. Even an infinitely large network
needs only a finite number of exchanged bits - about 400 in this case
- to synchronise.
 
\begin{figure} 
\centering 
\includegraphics[width=.6\textwidth]{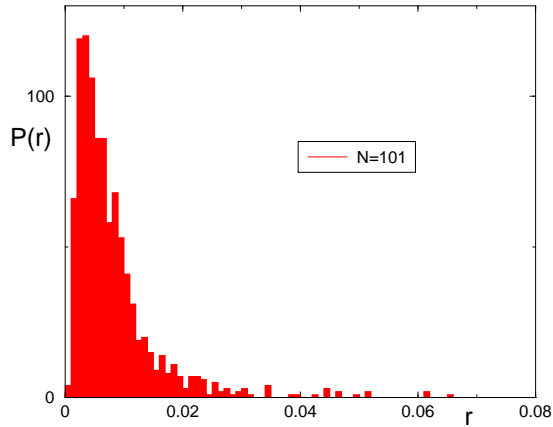} 
\caption[]{ 
  Distribution of the ratio of synchronisation time between networks A
  and B to the learning time of an attacker E.}
\label{rat} 
\end{figure}

If the communication continues after synchronisation, an opponent has
a chance to lock into the moving weights of $A$ and
$B$. Fig.\ref{rat} shows the distribution of the ratio between the
synchronisation time of $A$ and $B$ and the learning time of the
opponent. In the simulations for $N = 100$, this ratio never exceeded
the value $r=0.1$, and the average learning time is about 50000 time
steps, much larger than the synchronisation time. Hence, the two
partners can take their weights $\underline{w}^{A}_{i} (t) =
\underline{w}^{B}_{i} (t)$ at a time step $t$ where synchronisation
most probably occurred as a common secret key. Synchronisation of
neural networks can be used as a key exchange protocol over a public
channel.

Up to now it is not clear, yet, whether more advanced attacks will
finally break this exchange protocol. On the other side, there are
several possible extensions of the synchronisation mechanism where
tracking seems to be even harder \cite{KK}.
 
\section{Conclusions}

The dynamics of interacting neural networks has been studied in the
context of a simple model: the perceptron and its extensions. The
dynamics of these models can be calculated analytically; macroscopic
properties can be described by differential equations for a few order
parameters.

Several kinds of interaction processes have been studied. In all cases
the networks are trained by a set of examples which are generated by
different networks. This is the way networks interact. Some networks
are generating pairs of high dimensional input data and a
corresponding output signal and transmit this information to other
networks. The networks receiving this information are adapting their 
parameters -- their synaptic weights -- to each example. One question is
to what extend the networks exchanging information are approaching each other
in the high dimensional synaptic space.

The teacher/student scenario of a static networks generating the
examples in addition to an adaptive network being trained on these examples is
the case studied most. The question is: How well does the
student learn the rule which is producing the examples, how  can it
generalise? The analytic solutions for the perceptron show the number
of training examples has to be of the number of neurons to achieve
generalisation. For the optimal training algorithm the generalisation error 
decays not faster than the inverse power of training time.

If both teacher and student networks are more complex then new
phenomena are observed. For example, for a committee machine
specialisation occurs: For short times the student  relaxes fast to a
configuration for which it has achieved generalisation but it is still
in a symmetric state, it behaves like a simple perceptron. Only if the
number of examples is increased to a very large value the student network can
escape from this configuration and each member of the committee
specialises to its corresponding member in the teacher network.

A static teacher network can also generate a time series on which the
student network is trained. In addition to the overlap between teacher
and student parameters, one is interested in the prediction abilities
of the student after the interaction period. It turns out that
learning and prediction are not necessarily correlated. One finds
perceptrons which learn a chaotic sequence very well but cannot predict
it. On the other side, for quasiperiodic sequences it is difficult to
learn the weights of the teacher but it is simple to predict the
corresponding sequence.

A very general question about properties of prediction algorithms
leads to a perceptron interacting with itself.  It produces a rather
complex time series which yields 100 \% prediction error if the
same perceptron is trained on this sequence.  But even if a
different perceptron is trained on this sequence, it achieves some
knowledge about the teacher, hence its prediction error is larger than 50
\%.

A community of neural networks can exchange information and learn from
each other. It is shown how such a scenario can lead to successful
cooperation in the minority game -- a model for competing agents in a
closed market.

If two networks are exchanging information and learning from each
other, they can synchronise. That means, after some training time they
relax to a configuration with identical time dependent synaptic
weights (up to a common sign). The two networks keep diffusing on a
high dimensional hypersphere, but with identical weights. Neural
networks can synchronise by mutual learning.

This new phenomenon is applied to cryptography. It is shown how
multilayer networks with discrete weights synchronise after a few
hundred steps of interactions. However, a third network which is recording the
exchange of examples does not synchronise, at least during a short
period of time. Learning by mutual learning is fast, but learning by
listening is very slow. Hence two partners can agree on a common
secret key over a public channel. Any observer who knows all the
details of the algorithm and who knows all the training examples cannot
calculate the secret key. This is -- to my knowledge -- the first
public key exchange which is not based on number theory.  Future
research will show how well this new algorithm -- based on mutual
learning of neural networks -- can resist any advanced attacks which still
have to be invented.

Synchronisation of neural networks is an active subject of research in
neurobiology. Up to now it is -- to my knowledge -- unclear how
synchronisation develops and what is its function. Here the model
calculations point to a new direction: Two or several biological
networks can achieve a common time dependent state by learning
information exchanged between active partners. Any other network
receiving the same information without being able to influence the
 partners cannot lock into this time dependent common state.
Hence even in a fully connected network, parts of it can synchronise
by mutual learning, at the same time screening their synchronised state from
parts which learn only by listening.

In summary, this is  the first attempt to develop
a theory of interacting neural networks.  Several phenomena were
discovered from  simple models like the perceptron
or the parity machines. These phenomena were neither included into the
models from the beginning nor are obvious; they are a result of cooperative
behaviour of the synaptic weights and can only be understood from the 
analytical and numerical calculations.

As the different scenarios described in this overview show, the first
results of this theory of interacting adaptive systems may be relevant 
in the fields of  cooperative systems, nonlinear dynamics, time series
prediction, economic models, biological networks and cryptography. 

\vspace{1cm}

{\bf Acknowledgement} This overview is based on  enjoyable
collaborations with Ido Kanter, Richard Metzler and Michal
Rosen-Zvi. I thank Michael Biehl for suggestions on the manuscript. This work
has been supported by the German Israel Science Foundation (GIF), the
Minerva Center of the Bar Ilan University and the Max-Planck Institute
f\"ur Physik komplexer Systeme in Dresden.

\end{document}